\documentclass[aps,prd,showpacs]{revtex4}

\usepackage[colorlinks=true,urlcolor=blue,linkcolor=black,citecolor=blue]{hyperref}
\begin{document}
 
\title{ 
Fermion mass relations \\[.5ex]
in a supersymmetric SO(10) theory
}
 
\author{
Borut Bajc$^{(1)}$,  
Alejandra Melfo$^{(2,3)}$, 
Goran Senjanovi\'c$^{(3)}$ 
and Francesco Vissani$^{(4)}$}
\affiliation{$^{(1)}$ {\it J.\ Stefan Institute, 1001 Ljubljana, Slovenia}}
\affiliation{$^{(2)}$ {\it Centro de F\'{\i}sica Fundamental, 
Universidad de Los Andes, M\'erida, Venezuela}}
\affiliation{$^{(3)}${\it International Center for Theoretical Physics, 
Trieste, Italy}}
\affiliation{$^{(4)}${\it INFN, Laboratori Nazionali 
del Gran Sasso, Theory Group, Italy}} 
 
\begin{abstract} Neutrino and charged fermion masses provide important constraints 
on grand unified theories. We illustrate this by focusing on a renormalizable, 
supersymmetric SO(10) theory proposed long ago, that recently attracted great 
interest in view of its minimality. We show how the nature of the light Higgs, which 
depends on the GUT scale fields, gets reflected on the precise predictions for 
fermion masses and mixings. We exemplify this on the case of dominant Type II 
see-saw, which gets severely constrained and is likely to fail.

\end{abstract}


\maketitle
 
\section{Introduction}
 
The evidence for small neutrino mass finds its natural explanation in the see-saw 
mechanism \cite{seesaw}, which in turn points strongly towards the SO(10) grand 
unified theory \cite{Fritzsch:nn}. If so, one faces a clear task: construct the 
minimal theory, extract the low energy predictions and confront them with 
experiment. In practice, this means specifying the minimal Higgs sector that leads 
to a realistic theory. In turn, this depends on the realization of the see-saw, 
i.e.~whether one uses a ${\bf 16_H}$ or ${\bf \overline {126}_H}$ to give 
right-handed neutrinos mass (for a review see \cite{Senjanovic:2005sf}). 

In the former case, one either uses higher-dimensional operators or the radiatively 
induced (at two loops) $m_{\nu_R}$\cite{Witten:1979nr}. The radiative see-saw can 
work only for the non-supersymmetric theory or the strongly split supersymmetry 
\cite{Arkani-Hamed:2004fb,Giudice:2004tc}. Recently, the split-supersymmetric 
version was argued to be possibly realistic \cite{Bajc:2004hr} and the minimal 
version was constructed \cite{Bajc:2005aq}. One should recall that in SO(10) 
supersymmetry is not at all needed for gauge coupling unification. For the 
fermion mass and mixing situation in this case see the recent work 
\cite{Bajc:2005zf}.

The latter case--the tree-level $m_{\nu_R}$--was studied in recent years in great 
detail for the minimal version with ${\bf 10_H} $ and ${\bf \overline{126}_H}$ 
\cite{Lazarides:1980nt,Babu:1992ia}, and the further assumption of low-energy supersymmetry 
\cite{Aulakh:1982sw,Clark:1982ai}. In this theory, the charged fermion and neutrino 
masses stem from the same Yukawa interactions \cite{Lazarides:1980nt,Clark:1982ai,Babu:1992ia}, and 
thus one relates quark and lepton masses and mixings. For example, in the case of 
the type-II see-saw \cite{Lazarides:1980nt,Mohapatra:1980yp}, the large atmospheric 
mixing angle and the tiny corresponding $V_{cb}$ quark mixing are intimately related 
to the $b-\tau$ unification \cite{Bajc:2002iw}. Since with low-energy supersymmetry 
$m_b \sim m_\tau$ holds well at the GUT scale, this has boosted an in-depth study of 
the full three-generation situation \cite{Goh:2003sy,Bertolini:2005qb,Babu:2005ia}. Both type II 
and type I where analyzed at length and both scenarios appear to be realistic. A 
possible obstruction is the following: even if type II works beautifully for the 
structure of the neutrino mass matrix, it has a potential problem in predicting a 
large enough overall scale for neutrino masses. Thus it is indispensable that as 
careful as possible a study be performed in order to determine its fate.

The light Higgs composition \cite{Bajc:2004xe} depends explicitly on the nature of the GUT 
Higgs, and should be included in the analysis of fermion masses. This fact cannot be 
overemphasized. In other words, the predictions for fermion masses, especially 
$m_\nu$, depend not only on the choice of Yukawas, but equally on the nature of the 
light Higgs doublets, i.e.~the GUT Higgs. On the other hand, the minimal theory 
requires specifying the GUT Higgs; in this theory it is ${\bf 210_H}$ necessarily. 
The reason for this is the following: the light Higgs must be a mixture of ${\bf 
10_H}$ and ${\bf \overline{126}_H}$ bidoublets, for otherwise one would have single 
set of Yukawas, and this can only be achieved through a ${\bf 10_H \, 
\overline{126}_H \, 210_H}$ or ${\bf 126_H \, \overline{126}_H \, 210_H}$ couplings. 
Both terms leads also to a tiny VEV for the $(3,1,\overline{10})$ field contained in 
$\overline{126}_H$, providing directly a mass for the neutrinos (Type II see-saw). 
Thus it is clear that the decomposition of the light Higgs will enter in the 
structure of neutrino masses and mixings. This is actually true for both Type I and 
Type II see-saw, but we focus here for the sake of clarity and simplicity on the 
Type II case. Our findings indicate severe problems for this mechanism as the sole 
source of neutrino mass, as do the first attempts in this direction 
\cite{Aulakh:2005sq,charan,Aulakh:2005bd}.
 
\section{The supersymmetric SO(10) theory}

\begin{table}[t]
\begin{center}
\begin{tabular}{|l||c|}
\hline
$q_2 $&$ 4 x^2-3 x+1$\\ \hline
$q_3 $&$ 4 x^3-9 x^2+9 x -2 $\\ \hline
$p_2 $&$ (2 x-1)(x+1)$\\ \hline
$p_3 $&$ 12 x^3-17 x^2+10 x-1$\\ \hline
$p_4 $&$ (3 x-1)(x^3+5 x-1)$\\ \hline
$p_5 $&$ 9 x^5+20 x^4-32 x^3+21 x^2-7 x+1$\\ \hline
$p_6 $&$ 9 x^6-56 x^5+140 x^4-177 x^3+130 x^2-43 x+5$\\ \hline
$p_{10} $&$ 90 x^{10}-858 x^9+2009 x^8-3073 x^7+4479 x^6 -$\\ 
       &$  5018 x^5+3618 x^4 -1545 x^3+377 x^2-50 x +3$\\ \hline
\end{tabular}
\end{center}
\caption{\label{tab1}\em Polynomials of $x$ entering the expressions of the 
VEVs and the masses.}
\end{table}

Here we recall the definition of the SO(10) theory under consideration (for details, 
see \cite{Bajc:2004xe,Aulakh:2002zr,Fukuyama:2004ps}), discuss its free parameters, and derive the fermion 
mass relations. \subsection{Parameters of the theory} The superpotential of the 
model $W=W_{Y}+W_H$ is (see e.g., \cite{Aulakh:2003kg} for a more explicit notation):
\begin{equation}
\begin{array}{l}
\displaystyle
W_Y= {\bf 16_m ( {\it Y_{\bf 10}}\ 10 +  {\it Y_{\bf\overline{126}}}\ 
\overline{126}_H) 16_m},  \\
\displaystyle
W_H=\frac{m}{4!} {\bf 210_H}^2 + \frac{\lambda}{4!} {\bf 210_H}^3
+ \frac{M}{5!}  {\bf 126_H\ \overline{126}_H} 
+ \frac{\eta}{5!}  {\bf 126_H\ 210_H\ \overline{126}_H} + 
m_H^2 {\bf 10}^2 +\frac{1}{4!} {\bf 210_H\ 10}\ 
(\alpha {\bf 126_H}+\overline\alpha {\bf \overline{126}_H})
\label{supot}
\end{array}
\end{equation}
where $Y_{\bf 10}$ and $Y_{\bf\overline{126}}$ are the two Yukawa 
couplings of the theory. The rest of the superpotential describes the Higgs sector 
at high (GUT) scale.
  
Notice that in fact, this model has only one mass scale. The mass $m_H$ is fixed by 
the fine-tuning \cite{Aulakh:2003kg,Bajc:2004xe} condition: \begin{equation} m_H=m\times 
\frac{\alpha\overline\alpha}{2 \eta \lambda}\ \frac{p_{10}}{(x-1) p_3 p_5} 
\end{equation} that is needed to ensure a pair of light Higgs doublets $H_u$ and 
$H_d$ in the spectrum of the low energy supersymmetric standard model (MSSM). The 
parameter $x$ measures one of the vacuum expectation values (VEV)  in unities of 
$m/\lambda$ \cite{Aulakh:2003kg} (see below) and we denote with $p_i$, $i=1,2,3...$ certain 
polynomials of the variable $x$ listed in table \ref{tab1}. Also, the mass $M$ can 
be calculated in terms of $m$ and the other parameters due to the consistency 
condition: \begin{equation} M=m\times \frac{\eta}{\lambda}\ \frac{3-14 x + 15 
x^2-8x^3}{(x-1)^2} \end{equation} (that, if one likes so, can be thought as a 
replacement of $M$ in favor of $x$). Thus, any new scale, as the right-handed 
neutrino mass scale, has to arise by the mass $m$ times some adimensional quantity. 
One can then find the symmetry-breaking VEVs as functions of $x$ \cite{Aulakh:2003kg}
\begin{eqnarray}
\langle 1,1,1 \rangle_{210}=-\frac{m}{\lambda}\cdot \frac{x (1-5 x^2)}{(1-x)^2},
\ \langle 1,1,15 \rangle_{210}=
-\frac{m}{\lambda} \cdot \frac{(1 -2 x-x^2)}{(1-x)}, \nonumber  \\ 
\langle 1,3,1 \rangle_{210}=\frac{m}{\lambda} \cdot x,\
\langle 1,3,\overline{10} \rangle_{\overline{126}} \cdot \langle 1,3,10 \rangle_{126}= 
\frac{2 m^2}{\eta\lambda} \cdot
\frac{x (1-3 x) (1+x^2)}{(1-x)^2},
\label{vevs}
\end{eqnarray}

Let us introduce a convenient phase choice. Choosing the arguments
of the 4 Higgs fields as follows:
\begin{equation}
\begin{array}{l}
{\bf 210_H}\to - \mbox{arg}(m)/2 \\
{\bf 10_H}\to  \mbox{arg}(\eta+\lambda-\alpha-\overline\alpha-m)/2 \\
{\bf 126_H}\to  \mbox{arg}(\overline\alpha-\alpha-\eta-\lambda+2 m)/2 \\
{\bf \overline{126}_H}\to 
 \mbox{arg}(\alpha-\overline\alpha-\eta-\lambda + 2 m)/2 
\end{array}
\end{equation}
we see that (after fine-tuning) the Higgs sector has just 8 real parameters:
\begin{equation}
m,\ \alpha,\ \overline{\alpha},\ |\lambda|,\ |\eta|,\
\phi=\mbox{arg}(\lambda)=-\mbox{arg}(\eta),\  
x=\mbox{Re}(x)+ i \mbox{Im}(x).
\end{equation}
The parameter $x$ is known to be convenient to describe the VEVs and the 
masses of the particles; the dependences of observed fermion masses on the other 
parameters is rather simple, whereas the behavior in $x$ is usually non trivial 
\cite{Bajc:2004xe}. In this work we show that the $x$ is the single most important 
parameter for fermion masses, at the same level as the Yukawa couplings $Y_{\bf 10}$ 
and $Y_{\bf\overline{126}}$. (Actually we would dare to say that the only parameters 
with a comparable importance are the mass scale of the theory, $m$ and the unified 
gauge coupling).

\subsection{Fermion mass relations\label{secca}}

Here, we obtain the relations for charged and neutrino masses in this SO(10) theory.

Following \cite{Bajc:2004fj}, 
we obtain the relation for the up quark masses replacing the 
couplings $Y_{\bf 10}$ and $Y_{\bf\overline{126}}$ in $W_Y$--see 
eq.~(\ref{supot})--with the masses $M_d$ and $M_e$:
\begin{equation}
M_u=\frac{N_u}{N_d} \tan\beta \times [ M_d + \xi(x) (M_d-M_e) ]
\label{ch}
\end{equation}
where $N_u$ and $N_d$ are defined in the Appendix, $\tan\beta$ is the 
ratio of up-type and down-type VEVs of the MSSM Higgs fields, and
\begin{equation}
\xi(x) = \frac{x\ p_6}{2\ p_2\ p_5}
\label{xi}
\end{equation}
The polynomials of $x$ ($p_{i}$ $i=2,3,4,5...$) can be found in table 
\ref{tab1}. The derivation of this expression for $\xi$ can be done easily, using 
the explicit expressions for the light doublets reported in eq.~(\ref{ddd}) 
of the 
Appendix.

Let us pass to light (left) neutrinos. As shown in the Appendix, the seesaw formula 
reads (here, $v=174$ GeV in the scale of SU(2) breaking):
\begin{equation}
\begin{array}{ll}
\displaystyle
M_\nu=&
\displaystyle\frac{v^2}{ m}\times \frac{\sin^2\beta}{\cos\beta }\times 
\alpha \sqrt{\frac{|\lambda|}{|\eta|}} 
\frac{N_u^2}{N_d}
\times \left[ \frac{m_{I}}{v}  \, f_I(x) +\frac{m_{I\!I}}{v} \, f_{I\!I}(x) 
\right]\\[3ex]
&
m_{I}= M_e(M_d-M_e)^{-1} M_e-6\xi M_e+9 \xi^2 (M_d-M_e),\ \ \ 
m_{I\!I}= M_d-M_e, 
\end{array}
\label{nu1}
\end{equation}
where the functions contributing to Type I and Type II seesaw  are
\begin{equation}
f_I(x)=\frac{2 p_2\ p_5}{p_3\ \sigma}, \ \ 
f_{I\!I}(x)= 
\frac{(x-1)\ (4x-1)\ p_3\ q_3^2\ \sigma}{2\ x\ p_2\  p_5\ q_2}
\label{nu3}
\end{equation}
where the polynomials of $x$ ($p_{i}$ and $q_i$, $i=2,3,4,5...$) are given 
in the table, and $\sigma(x)$ in eq.~(\ref{sigma}). These are the equations needed 
when addressing fermion mass fitting in the model under consideration. 

\section{Constraints from fermion masses}

\subsection{Information on $x$ parameter from charged fermions}

In order to analyze fermion masses, a convenient flavor basis is the one where $M_d$ 
is diagonal (with positive entries on the diagonal), and therefore $M_u$ is rotated 
by the conventional CKM mixing matrix \cite{Bajc:2004fj}. Note that in this basis, both 
matrices $M_u$ and $M_e$ should include a number of phases, that are however 
unobservable at low energies.

An important quantity that we can determine from the fit of 
charged fermion masses is the relative amount of $M_d$ and $M_e$
entering in eq.~(\ref{ch}), namely:
\begin{equation}
R=|1+1/\xi|
\label{erre}
\end{equation}
This quantity is a function of $\xi$ only, and therefore, gives 
information on the parameter $x$. The behavior of $R$ in the $x$-plane is quite 
easy to understand; in seven points (the roots of $p_2$ and $p_5$), $R$ goes to 
zero; in other seven points ($x=0$ and the roots of $p_6$, e.g., $x_\pm^{(1)}\sim 
0.3 (1\pm i/10)$) $R$ goes to infinity; at infinity, $R$ goes to 5.

Fit of the data suggest a value of $R\in [2,4]$, as for example in \cite{Bertolini:2005qb}. 
The allowed range of $R$ however can depend on details of the fitting procedure. 
Analytical approximate arguments suggest that $R$ is closely connected with the GUT 
value of the mass ratio $m_\mu/m_s$, which is consistent with the above given range. 
For our illustrative purposes, it will be sufficient to consider the lower bound
\begin{equation} 
R > 1
\end{equation}
which stems from trace identities of (\ref{ch}) valid in the real case 
\cite{Malinsky}. As we discuss below, this very conservative limit is enough to 
endanger the viability of models with Type II see-saw dominant.
 
\subsection{How to enhance the neutrino mass scale}

The typical scale of neutrino masses is $v^2/m$. However, a value of $m\sim M_{GUT}$ 
as the mass scale suggested by gauge coupling unification or proton decay bounds 
would lead to unacceptably small neutrino masses. Thus, we are interested in having 
a neutrino mass scale sensibly higher than $v^2/M_{GUT}$. There are only a few 
possibilities to achieve that. A first possibility is $|\eta|\ll |\lambda|$; a 
second one is to consider certain points in the $x$-plane, where the function 
$f_{I\!I}$ is large.

The first possibility is hardly viable. A small $\eta$ will increase the 
contribution of the (2,2,15) and (2,2,10) doublets to the light Higgs, with the 
result that the coefficient $N_u^2/N_d$ diminishes accordingly, as can be seen from 
the explicit expression in the Appendix.  Indeed, the left-handed triplet VEV is 
induced by the term $ \eta\, {\bf 210_H \, \overline{126}_H\, 126_H}$, so it 
disappears in the limit $\eta \to 0$.  On the other hand, diminishing $\eta$ has the 
effect of diminishing the masses of a large number of supermultiplets from 
the ${\bf 126_H}$ and ${\bf \overline{126}_H}$ fields. Although a careful study of the RGE equations 
taking into account threshold effects is in order here, it will be extremely 
difficult to satisfy unification constrains, or even worse, to keep the Landau pole 
acceptably away from $M_{GUT}$.  In the following, we will consider all couplings of 
the same order.

The second possibility requires a detailed discussion. Consider first the limit 
$x\to \infty$. It is easy to see from (\ref{vevs}) that in this limit, the GUT scale 
Higgs field VEVs become hierarchical, indicating the onset of the multiple-step 
symmetry breaking regime. The possibility of having intermediate scales has been 
studied and discarded in \cite{Bajc:2004xe} through RGE analysis. The point is that the 
single step breaking works very well, and resists alternative solutions 
\cite{unification}. Simply stated, in the multiple-step case too many particles 
become light, spoiling unification. We will therefore keep $x$ safely small.

The type I seesaw contribution increases when $f_I(x)$ is large, namely where 
$\sigma\to 0$ ($x=0,1/3,\pm i$) and where $p_3\to 0$ ($x\sim 0.12,0.64\pm i 0.51$). 
Note that in the second set of points the component of the doublet 
$v_d^{\overline{126}}\in {\bf\overline{126}}$ is large, so that at the same time 
$N_d$ goes to zero.

The type II seesaw increases where $x\to 0$, where $q_2\to 0$ ($x = (3\pm i 
\sqrt{7})/8$), where $p_2\to 0 $ ($x\sim 1/2$ and $x\sim -1$), and where $p_5\to 0$ 
($x\sim -3.5,.26\pm i .39,.36\pm i.13$). In the last 7 points, the component of the 
doublet $v_d^{\overline{126}}$ (and also $v_d^{10}$) is very small. Note that when 
$v_d^{\overline{126}}$ becomes small, $Y_{\bf\overline{126}}$ increases in order to 
keep $M_d-M_e$ fixed; in this way, one can put a bound from perturbativity. This 
however has a clear physical meaning: the type II seesaw contribution is maximal 
when $Y_{\bf\overline{126}}$ is larger. The summary of the discussion is in table 
\ref{tab2}.

\begin{table}
\begin{center}
\begin{tabular}{|l|l|c|c|}
\hline
$x$           & $R$ & seesaw & remarks \\ \hline
1/3     & 3 & $I$ & flipped SU(5) \\
$\pm i$ & 3.14 &  $I$ & SM$\times$U(1) \\
$p_3(x)\to 0$ & $\infty$ & $I$ & small $Y_{\bf\overline{126}}$ \\ 
0       &  $\infty$ & $I$+$I\!I$ & left-right \\
$(3 \pm i\sqrt{7})/8$ & 0.62 &  $I\!I$ & light tripl. \\ 
$p_2(x)\to 0$ & 1 & $I\!I$ & large $Y_{\bf\overline{126}}$ \\ 
$p_5(x)\to 0$ & 1 & $I\!I$ & large $Y_{\bf\overline{126}}$ \\ \hline
\end{tabular}
\end{center}
\caption{\label{tab2}\em 
$1^{st}$ column, special points of the $x$ plane where
neutrino mass scale is enhanced;
$2^{nd}$ column, values of $R$ in these points;
$3^{rd}$ column, seesaw type;
$4^{th}$ column, intermediate symmetry breaking or other 
characteristic features.}
\end{table}

\subsection{Are fermion mass relations compatible with data?}

Here, we would like to concentrate the discussion on pure Type II seesaw and point 
out that the conditions on $x$ from charged fermion masses and the one from the mass 
scale of neutrinos are not automatically consistent (however we note on passing 
that, although reasonable fits with this Type I seesaw have been obtained 
\cite{Matsuda:2001bg,Babu:2005ia}, the result is not fully satisfactory when compared with 
present neutrino data, see e.g., \cite{Strumia:2005tc,Fogli:2005cq}).

From table \ref{tab2}, we see that, excluding the cases where the Yukawa couplings 
become large, there are in fact only two points where the Type II mass can be 
sufficiently large, $x \sim 0$ and $x \sim (3 \pm i \sqrt{7})/8$. It is easy to 
understand why: the Type II neutrino mass is given by the relation
\begin{equation} M_\nu\propto m_{II} \propto Y_{\bf\overline{126}} \times 
\frac{ (\alpha  v_u^{\bf 10} + \sqrt{6} \eta
v_u^{\bf\overline{126}} )  v_u^{\bf 210}}{m_{tripl}}
\end{equation}
where the triplet mass is
\begin{equation}
m_{tripl} = \frac{\eta}{\lambda} m \frac{x q_2}{(x -1)^2} 
\label{eq10}
\end{equation}
Thus the selected points are simply those where the triplet becomes light. 
With $x \sim 0$, the symmetry breaks down in two steps, with and intermediate 
left-right group \cite{leftright}, and as we said the possibility of intermediate 
scales is excluded by the unification constraints.

The other points, roots of $q_2$, give a mass spectrum with all particles around 
$M_{GUT}$, with the sole exception of the triplet.  A careful study in the vicinity 
of these singular points is in order. However, we can already see that it will be 
very difficult to fit the fermion spectrum, given that $R < 1$ at those points (see 
table \ref{tab2}).

Unification constraints can also be a problem. With no other light particles to 
compensate for the effects of a triplet around $10^{14}$ GeV, it is very unlikely to 
work. This issue should be addressed in a full two-loop calculation, including 
threshold effects. We suspect however that fermion mass fitting can be sufficient to 
rule out the model.

\section{Discussion}

This work addresses an important issue faced when looking for the predictions of a 
well-defined simple GUT such as SO(10): the dependence of the fermionic mass spectra 
on the nature if the light Higgs. The nature, i.e.~the decomposition of the light 
Higgs among the plethora of such fields present in the theory, is a direct 
consequence of the GUT Higgs structure, hence this way one probes the physics 
apparently out of each. We exemplify this on the minimal SUSY SO(10) theory with 
${\bf 10_H}$ and ${\bf \overline{126}_H}$ recently studied at length and in this 
context we emphasize the difference between fermion masses as predicted by a 
``generic'' model with a ${\bf {10_H}}$ and a ${\bf \overline{126}_H}$, and those 
predicted in the supersymmetric SO(10) theory with ${\bf 210_H}$ defined above. 

We should recall here that the generic case, i.e. independently of how many 
${\bf 210_H}$ fields are used or whether one adds also ${\bf 45_H}$ and/or 
${\bf 54_H}$ fields, matter parity (equivalent to R-parity) is a gauge symmetry 
\cite{rparity} and it remains unbroken at all energies \cite{lrsusy}. This 
guarantees the stability of the lightest supersymmetric partner, an ideal 
candidate for the dark matter. The theory furthermore predicts the hierarchical 
neutrino mass spectrum and an appreciable leptonic $1-3$ mixing: 
$\left|U_{e3}\right| >0.1$, and thus can be tested and possibly even ruled out. 

It is important though to work out precisely the predictions of the minimal theory 
\cite{Aulakh:2003kg} which uses the single ${\bf 210_H}$ as the GUT Higgs. The essential point in 
this case is that the light Higgs is not an arbitrary combination of the plethora of 
the original fields, but it is determined through the same interactions that decide 
on the spectra and couplings of the heavy fields \cite{Bajc:2004xe}. This further 
restrains the fermion mass relations as indicated in the previous studies 
\cite{Aulakh:2005sq,charan,Aulakh:2005bd}, prompting us to address this issue more carefully. We 
focused the analysis on Type II see-saw and found that he constraints are severe. In 
particular, the type II seesaw dominates for the sufficiently light SU(2)$_L$ 
triplet in ${\bf \overline{126}_H}$. However, in this overconstrained theory, its mass is 
set by the same parameters that determine the light Higgs, and force it to be large. 
To paraphrase the Poet: We came to praise 
the model, not to bury it... but 
we fear we are on the verge of doing the latter.
Still, we are convinced that it is essential
to check with the greatest care whether such a
pessimistic view is correct:
Not only in the hope to find a loophole in our argument,
but at least to demonstrate that well-defined grand 
unified models are positively testable.

\section{Acknowledgments} Some of the results of this work were already presented by 
G.S. in the plenary session of PASCOS 05 and by A.M. at the 2005 Gran Sasso Summer 
Institute (a good part of this work took place there). G.S.~is grateful to the 
organizers of the PASCOS 05 conference, especially J.E.~Kim, for a kind invitation, 
warm hospitality and an extremely well-organized conference. We thank Zurab 
Berezhiani for a great meeting and hospitality. 
The work of G.S.\ was supported in 
part by European Commission under the RTN contract MRTN-CT-2004-503369; the work of 
B.B.\ by the Ministry of Education, Science and Sport of the Republic of Slovenia: 
the work of A.M.\ by CDCHT-ULA project No.\ C-1244-04-05-B
and FONACIT F-2002000426.  We thank Charan Aulakh, 
Michal 
Malinsk\'y and Miha Nemev\v{s}ek for discussion and correspondence.

\appendix\section{Derivation of the formula for neutrino masses in Sec.~\ref{secca}} 

We will discuss piece by piece the lagrangian of massive neutrinos in our model. 
Suppressing flavor indices, and denoting by $\nu$ the left neutrino and by $\nu^c$ 
the (conjugate of the) right handed neutrino, we have:
\begin{equation}
{\cal L}=
\frac{1}{2} \nu^c (c_R Y_{\bf\overline{126}} \langle 1,3,\overline{10}\rangle) \nu^c +
\nu (c Y_{\bf 10} v_u^{10}-3 c' Y_{\bf\overline{126}} v_u^{\overline{126}}) \nu^c +
\frac{1}{2} \nu (c_L Y_{\bf\overline{126}} \langle 3,1,10\rangle) \nu + h.
\end{equation}
where the first and the third (resp., the second) bracketed term are Majorana 
(resp., Dirac) masses. The numerical coefficients $c,c',c_L,c_R$ are the 
Clebsch-Gordan coefficients of the spinor-spinor-Higgs couplings. The VEVs of 
positive hypercharge doublets are denoted as $v_u^{10}$ and $v_u^{\overline{126}}$, 
whereas $\langle 1,3,\overline{10}\rangle$ and $\langle 3,1,10\rangle$ are the 
singlet and triplet VEV. The presence of the Yukawa couplings $Y_{\bf 10}$ and 
$Y_{\bf\overline{126}}$ is evident, when we note that $v_u^{10}\in {\bf 10}$ and 
$\langle 1,3,\overline{10}\rangle , v_u^{\overline{126}},\langle 3,1,10\rangle\in 
{\bf \overline{126}}$. Any non-trivial dependence on the parameters of $W_{H}$ is 
buried in these 4 VEVs, as it will be clear from their explicit expressions:

\begin{itemize}

\item 

We begin with the singlet:
\begin{equation}
\langle 1,3,\overline{10}\rangle=\frac{m}{\sqrt{|\eta \lambda|}} \times \sigma(x), \ \ 
\sigma=\sqrt{\frac{2 x (1-3x)(1+x^2)}{(1-x)^2}}
\label{sigma}
\end{equation}
which shows that neutrino masses are small only for certain values of $x$ 
\cite{Aulakh:2003kg}.

\item 

Passing to the doublets, let us consider the positive hypercharge Higgs $H_u$ of the 
MSSM (analogous considerations apply to $H_d$). Its VEV is given by a sum of the 4 
component VEVs $v_u^{10},v_u^{\overline{126}},v_u^{126},v_u^{210}$, contained in the 
${\bf 10_H}$, ${\bf \overline{126}_H}$, ${\bf 126_H}$, ${\bf 210_H}$, respectively. 
We factorize the dependence on $v\approx 174$~GeV and on $\tan\beta$ in a
natural  manner,
\begin{equation}
v_u^{10}=v\sin\beta\times N_u\ \xi_u^{10},\ 
v_u^{\overline{126}}=v\sin\beta\times N_u\ \xi_u^{\overline{126}},\ \   \mbox{etc}
\end{equation}
where the proper normalization is ensured by
\begin{equation}
N_u=1/\sqrt{|\xi_u^{10}|^2+|\xi_u^{\overline{126}}|^2+|\xi_u^{126}|^2+|\xi_u^{210}|^2}.
\end{equation}
An analytical expression of the 4 VEVs is then 
obtained using \cite{Bajc:2004xe}:
\begin{equation}
\begin{array}{cccc}
\xi_u^{10}=\frac{2 p_5}{x-1} & 
\xi_u^{\overline{126}}=-\sqrt{6} \frac{\alpha}{\eta} p_4 & 
\xi_u^{126}=-\sqrt{6} \frac{\overline{\alpha}}{\eta} p_2 p_5/p_3 & 
\xi_u^{210}=-\frac{{\alpha}}{\sqrt{|\eta\lambda|}} q_3 \sigma \\
\xi_d^{10}=\frac{2 p_5}{x-1} & 
\xi_d^{\overline{126}}=-\sqrt{6} \frac{\alpha}{\eta} p_2 p_5/p_3 & 
\xi_d^{126}=-\sqrt{6} \frac{\overline{\alpha}}{\eta} p_4 &
\xi_d^{210}=\frac{{\overline{\alpha}}}{\sqrt{|\eta\lambda|}} q_3 \sigma 
\end{array}
\label{ddd}
\end{equation}
where we give the analogous and 
obvious definitions of $N_d$ and 
of $v_d^{10},v_d^{\overline{126}}...$ (with $\sin\beta\to \cos\beta$).

\item Finally, the triplet VEV is
\begin{equation}
\langle 3,1,10\rangle=\frac{(\alpha v_u^{10} + \sqrt{6} \eta v_u^{\overline{126}})
v_u^{210}}{m_{tripl}}= -\frac{v^2\sin^2\beta N_u^2}{m}\times 
\frac{\alpha^2 \lambda }{\eta \sqrt{|\eta\lambda|}}\times 
\frac{(x-1)(4 x-1) q_3^2\sigma}{x q_2}
\end{equation}
the first equality shows that this is an induced VEV, involving the couplings 
$\alpha$ and $\eta$ \cite{Aulakh:2003kg}; the Clebsh-Gordan coefficients are taken from 
\cite{Aulakh:2005bd}. The last equality is obtained by using eq.~(\ref{ddd}) and the 
expression of the triplet mass in eq.~(\ref{eq10}), $m_{tripl}=m\ \eta/\lambda\ x 
q_2/(x-1)^2$ \cite{Aulakh:2003kg}. The VEV is larger where the mass of the triplet is small, 
namely at $x\sim 0$ (note that $\sigma= {\cal O}(\sqrt{x})$ is also small there) and 
where $q_2$ is small, namely in $x=(3\pm i\sqrt{7})/8$~\cite{Aulakh:2003kg}.

\end{itemize}

There are still two steps to do:

\begin{itemize}

\item We take the Clebsh-Gordan 
coefficients calculated in \cite{charan,Aulakh:2005bd}:
\begin{equation}
c_R=16,\  c_L=-16,\ c=2\sqrt{2},\ c'=-i 4\sqrt{2/3}
\end{equation}

\item We can replace as suggested in \cite{Bajc:2004fj}, 
the couplings $Y_{\bf 10}$ and $Y_{\bf\overline{126}}$ with 
the masses $M_d$ and $M_e$, 
using in particular the relation:
\begin{equation}
Y_{\bf\overline{126}}=\frac{M_d-M_e}{4 c' v_d^{\overline{126}}}=
\frac{M_d-M_e}{32 i v \cos\beta}\times 
\frac{\eta\ p_3}{\alpha\ N_d\ p_2 p_5}
\end{equation}
and a relation for the Dirac mass, similar to the one for $M_u$.
Note as a curiosity that the explicit factors 
16 appearing in Majorana masses through $c_L,c_R$
simplify with the factor 32  
in previous denominator.

\end{itemize}

In summary: we replace the Yukawa couplings with $M_d$ and $M_e$, plug in the 
Lagrangian the various VEVs and constants listed above, use the well-known seesaw 
formula to integrate away right-handed neutrinos, neglect the irrelevant overall 
phase, and obtain in this way equations (\ref{nu1})-(\ref{nu3}).


\begin{thebibliography}{99}

\bibitem{seesaw}
P.~Minkowski,
Phys.\ Lett.\ B {\bf 67} (1977) 421; 
T.~Yanagida, proceedings of the {\em Workshop on Unified Theories 
and Baryon Number in the Universe}, Tsukuba, 1979, eds. 
A. Sawada, A. Sugamoto; 
S.~Glashow, in {\em Cargese 1979, Proceedings, Quarks and Leptons}
(1979) ;
M.~Gell-Mann, P.~Ramond, R.~Slansky, proceedings of the
{\em Supergravity Stony Brook Workshop}, New York, 1979, 
eds. P. Van Niewenhuizen, D. Freeman; 
R.~Mohapatra, G.~Senjanovi\' c,
Phys.Rev.Lett. {\bf 44} (1980) 912.

\bibitem{Fritzsch:nn}
H. Georgi, {\em In Coral Gables 1979 Proceeding, Theory and experiments
in high energy physics,} New York 1975, 329 and
H.~Fritzsch and P.~Minkowski,
Annals Phys.\  {\bf 93} (1975) 193.

\bibitem{Senjanovic:2005sf}
  G.~Senjanovi\' c, 
Talk given at SEESAW25: International Conference on the Seesaw 
Mechanism and the Neutrino Mass, Paris, France, 10-11 Jun 2004.
Published in *Paris 2004, Seesaw 25* 45-64; 
  arXiv:hep-ph/0501244.

  

\bibitem{Witten:1979nr}
E.~Witten,
Phys.\ Lett.\ B {\bf 91} (1980) 81.

\bibitem{Arkani-Hamed:2004fb}
N.~Arkani-Hamed and S.~Dimopoulos,
arXiv:hep-th/0405159.

\bibitem{Giudice:2004tc}
G.~F.~Giudice and A.~Romanino,
Nucl.\ Phys.\ B {\bf 699} (2004) 65
[arXiv:hep-ph/0406088].
N.~Arkani-Hamed, S.~Dimopoulos, G.~F.~Giudice and A.~Romanino,
arXiv:hep-ph/0409232.


\bibitem{Bajc:2004hr}
  B.~Bajc and G.~Senjanovi\'c,
  Phys.\ Lett.\ B {\bf 610} (2005) 80
  [arXiv:hep-ph/0411193].

\bibitem{Bajc:2005aq}
  B.~Bajc and G.~Senjanovi\'c,
  arXiv:hep-ph/0507169 (PRL in press).


\bibitem{Bajc:2005zf}
  B.~Bajc, A.~Melfo, G.~Senjanovi\' c and F.~Vissani,
  arXiv:hep-ph/0510139.

\bibitem{Lazarides:1980nt}
  G.~ Lazarides, Q.~Shafi and C.~Wetterich,
  Nucl.\ Phys.\ B {\bf 181} (1981) 287.



\bibitem{Babu:1992ia}
K.~S.~Babu and R.~N.~Mohapatra,
  Phys.\ Rev.\ Lett.\  {\bf 70} (1993) 2845
  [arXiv:hep-ph/9209215].

 
\bibitem{Aulakh:1982sw}
  C.~S.~Aulakh and R.~N.~Mohapatra,
  Phys.\ Rev.\ D {\bf 28} (1983) 217.



\bibitem{Clark:1982ai}
T.~E.~Clark, T.~K.~Kuo and N.~Nakagawa,
  Phys.\ Lett.\ B {\bf 115}, 26 (1982).

\bibitem{Mohapatra:1980yp}
  R.~N.~Mohapatra and G.~Senjanovi\' c,
  Phys.\ Rev.\ D {\bf 23} (1981) 165.

\bibitem{Bajc:2002iw}
B.~Bajc, G.~Senjanovi\'c and F.~Vissani,
hep-ph/0110310  and 
  Phys.\ Rev.\ Lett.\  {\bf 90} (2003) 051802
  [arXiv:hep-ph/0210207].

 
\bibitem{Goh:2003sy}
 H.~S.~Goh, R.~N.~Mohapatra and S.~P.~Ng,
  Phys.\ Lett.\ B {\bf 570} (2003) 215
  [arXiv:hep-ph/0303055] and 
  Phys.\ Rev.\ D {\bf 68} (2003) 115008
  [arXiv:hep-ph/0308197].

\bibitem{Bertolini:2005qb}
S.~Bertolini and M.~Malinsky,
  arXiv:hep-ph/0504241.

\bibitem{Babu:2005ia}
K.~S.~Babu and C.~Macesanu,
  arXiv:hep-ph/0505200.

\bibitem{Bajc:2004xe}
B.~Bajc, A.~Melfo, G.~Senjanovi\' c and F.~Vissani,
  Phys.\ Rev.\ D {\bf 70} (2004) 035007
  [arXiv:hep-ph/0402122].
 
 \bibitem{Aulakh:2005sq}
C.~S.~Aulakh,
  hep-ph/0501025;

\bibitem{charan}
C.~S.~Aulakh and A.~Girdhar,
  Nucl.\ Phys.\ B {\bf 711} (2005) 275.

\bibitem{Aulakh:2005bd}
C.~S.~Aulakh,
  arXiv:hep-ph/0506291.

\bibitem{Aulakh:2002zr}
C.~S.~Aulakh and A.~Girdhar,
  Int.\ J.\ Mod.\ Phys.\ A {\bf 20} (2005) 865
  [arXiv:hep-ph/0204097].

\bibitem{Fukuyama:2004ps}
T.~Fukuyama, A.~Ilakovac, T.~Kikuchi, S.~Meljanac and N.~Okada,
  J.\ Math.\ Phys.\  {\bf 46}, 033505 (2005).

\bibitem{Aulakh:2003kg}
C.~S.~Aulakh, B.~Bajc, A.~Melfo, G.~Senjanovi\'c and F.~Vissani,
  Phys.\ Lett.\ B {\bf 588} (2004) 196
  [arXiv:hep-ph/0306242].
 
\bibitem{Bajc:2004fj}
B.~Bajc, G.~Senjanovi\'c and F.~Vissani,
  Phys.\ Rev.\ D {\bf 70} (2004) 093002
  [arXiv:hep-ph/0402140].

\bibitem{Malinsky} M. Malinsky, SISSA Ph.D.~Thesis (Oct.~2005). 

 \bibitem{unification}
S.~Dimopoulos, S.~Raby, F.~Wilczek,
Phys.\ Rev.\ D {\bf 24} (1981) 1681.
L.E.~Ib\'a\~nez, G.G.~Ross,
Phys.\ Lett.\ B {\bf 105} (1981) 439.
M.B.~Einhorn, D.R.~Jones,
Nucl.\ Phys.\ B {\bf 196} (1982) 475.
W.~Marciano, G.~Senjanovi\' c,
Phys.Rev.D {\bf 25} (1982) 3092.

\bibitem{Matsuda:2001bg}
K.~Matsuda, Y.~Koide, T.~Fukuyama and H.~Nishiura,
  Phys.\ Rev.\ D {\bf 65} (2002) 033008
  [Erratum-ibid.\ D {\bf 65} (2002) 079904]
  [arXiv:hep-ph/0108202].


\bibitem{Strumia:2005tc} 
A.~Strumia and F.~Vissani,
Nucl.\ Phys.\ B {\bf 726} (2005) 294.

\bibitem{Fogli:2005cq}
G.~L.~Fogli, E.~Lisi, A.~Marrone and A.~Palazzo,
  arXiv:hep-ph/0506083.

\bibitem{leftright}
J.~C.~Pati and A.~Salam,
Phys.\ Rev.\ D {\bf 10} (1974) 275.
R.~N.~Mohapatra and J.~C.~Pati,
Phys.\ Rev.\ D {\bf 11} (1975) 2558.
G.~Senjanovi\'c and R.~N.~Mohapatra,
Phys.\ Rev.\ D {\bf 12} (1975) 1502.
G.~Senjanovi\'c,
Nucl.\ Phys.\ B {\bf 153} (1979) 334.
 

\bibitem{rparity}
R.~N.~Mohapatra,
{\em Phys.\ Rev.\ } {\bf D 34}, 3457 (1986).
A.~Font, L.~E.~Ib\'a\~nez and F.~Quevedo,
{\em Phys.\ Lett.\ } {\bf B228}, 79 (1989).
S.~P.~Martin,
{\em Phys.\ Rev.\ } {\bf D46}, 2769 (1992).

\bibitem{lrsusy}
C.S.~Aulakh, K.~Benakli, G.~Senjanovi\'c,
Phys.\ Rev.\ Lett.\  {\bf 79} (1997) 2188.
C.~S.~Aulakh, A.~Melfo and G.~Senjanovi\'c,
Phys.\ Rev.\ D {\bf 57}, 4174 (1998).
C.~S.~Aulakh, A.~Melfo, A.~Ra\v{s}in and G.~Senjanovi\'c,
Phys.\ Lett.\ B {\bf 459} (1999) 557.
C.~S.~Aulakh, B.~Bajc, A.~Melfo, A.~Ra\v{s}in and G.~Senjanovi\'c,
Nucl.\ Phys.\ B {\bf 597} (2001) 89.

\end{thebibliography}
\end{document}